\documentclass[letterpaper, 10 pt, conference]{ieeeconf}  

\IEEEoverridecommandlockouts                              

\usepackage{subcaption}
\usepackage{color}

\usepackage{cite}
\usepackage{algorithm}
\usepackage{algpseudocode}
\usepackage[algo2e]{algorithm2e}
\usepackage{amsmath}
\usepackage{multirow}
\usepackage{multicol}   
\usepackage[justification=centering]{caption}
 
\newtheorem{theorem}{Theorem}
\newtheorem{lemma}[theorem]{Lemma}

\usepackage{graphicx}
\usepackage[bottom]{footmisc}

\usepackage{caption}
\usepackage{subcaption}

\usepackage{epstopdf,float,amsfonts}
\usepackage{epsfig}

\usepackage{todonotes} 

 \title{\Large \bf Resilient Communication Scheme for Distributed Decision of Interconnecting Networks of Microgrids}
\author{Thanh Long Vu,~\IEEEmembership{Member,~IEEE,} Sayak Mukherjee,~\IEEEmembership{Member,~IEEE}, Veronica Adetola,~\IEEEmembership{Senior Member,~IEEE} 
\thanks{Authors are with the Pacific Northwest National Laboratory, Richland, WA, USA.
Corresponding author: Thanh Long Vu (email: thanhlong.vu@pnnl.gov).}}

\begin{document}

\maketitle
\begin{abstract}
   Networking of microgrids can provide the operational flexibility needed for the increasing number of DERs deployed at the distribution level and supporting end-use demand when there is loss of the bulk power system. But, networked microgrids are vulnerable to cyber-physical attacks and faults due to the complex interconnections. As such, it is necessary to design resilient control systems to support the operations of networked microgrids in responses to cyber-physical attacks and faults. This paper introduces a resilient communication scheme for interconnecting multiple microgrids to support critical demand, in which the interconnection decision can be made distributedly by each microgrid controller even in the presence of cyberattacks to some communication links or microgrid controllers. This scheme blends a randomized  peer-to-peer communication network for exchanging information among controllers and resilient consensus algorithms for achieving reliable interconnection agreement. The network of 6 microgrids divided from a modified 123-node test distribution feeder is used to demonstrate the effectiveness of the proposed resilient communication scheme. 
\end{abstract}


\section{Introduction}

Distributed energy resources (DERs), such as rooftop PV and battery energy storage systems (BESS), are increasingly deployed in electric distribution systems to support utility operations, as well as end-use customers. As points of aggregation for DERs, microgrids have been considered as the enabling technology to support the integration of DERs. When there are increasing number of microgrids, it is possible to coordinate the operations of networks of microgrids \cite{7922501}. Interconnecting multiple microgrids can enhance their operational flexibility, e.g., the strong system with redundant supply can support the other weaker systems to provide better services to customers.

However, an interconnection of microgrids is also vulnerable to attacks and faults due to the complex interactions. 
Various types of cyberattacks to control systems have been reported in  academic study and industrial vulnerability assessment, e.g., denial-of-service attacks \cite{yuan2013resilient}, false data-injection attacks \cite{bai2017data}, replay attacks \cite{zhu2013performance}, covert attacks \cite{ barboni2020detection}, etc.  Therefore, to support the reliable operations of networked microgrids, it is necessary to take the resilience-to-attack into consideration when designing their control systems. 

In this paper, a resilient-to-cyberattack communication scheme is introduced to support the interconnection decision of networked microgrids in the presence of cyberattacks. To reduce the risk of one-point-failure, it is assumed that there is no centralized controller coordinating all the microgrids. The removal of centralized controller is also necessary in a mixed ownership environment where each microgrid belongs to either utility and non-utility, and hence, it is not authorized to collect information from all microgrids by a centralized controller. Recently, \cite{8618618} presented a distributed architecture using the Open Field Message Bus (OpenFMB) that enables peer-to-peer (p2p) communication at the application layer to be implemented. Using such a p2p communication network, the microgrid controllers exchange information to make the interconnection decision in a distributed manner. To make it difficult for an attacker to be successful in targeting the communication links, we propose a strategy whereby a communication agent randomly generate the p2p communication graph (i.e., graph of who-talk-with-who), without getting information from microgrids. Then, by using resilient consensus algorithms, the microgrid controllers can reach interconnection agreement even in the presence of cyberattacks to some controllers.

Designing advanced controllers for resilient  networked microgrids  have been investigated. In \cite{selfhealingNM}, networked microgrids (MG) were investigated for self-healing purpose when a generation deficiency or fault happens in a MG. Here, the MGs are connected through a common point of coupling, and both communication and physical network topologies are fixed. A framework for assembling of networked microgrids  was introduced in \cite{schneider2022framework}; yet the communication network was also fixed to an all-to-all topology. Our work advances the current state-of-the-art by considering cyberattack to the communication links and controllers and leveraging the flexibility of p2p communication network.

The novelty and contributions of this paper include:
\begin{itemize}
    \item A resilient p2p communication framework is introduced for the distributed decision of interconnecting networks of microgrids in the presence of cyberattacks to communication links and microgrid controllers.
    \item Resilient consensus algorithm is investigated to robustly evaluate the networked microgrid's interconnection criteria when there are attacked controllers. In this paper, the criteria is based on the total supply and critical demand of all participating microgrids.
    \item The key to success of the resilient consensus algorithm is to have a communication graph with enough vertex-connectivity. Accordingly, two algorithms are introduced for  generating a communication graph with a given vertex-connectivity, in both cases when the cyberattacks to communication links are known or unknown.
\end{itemize}

The remainder of this paper is organized as follows. In Section \ref{system-problem}, the physical and cyber architecture supporting distributed decisions in networked microgrids is presented. Also, the interconnection criteria and problem statement are introduced. Section \ref{algorithms} provides the graph algorithms for generating a p2p communication network and the resilient consensus algorithms for the interconnection decision making. Numerical demonstration on a co-simulation platform is presented in Section \ref{demonstration}, followed by conclusions.

 \textit{Graph theory preliminaries:} The undirected graph $\mathcal{G} = (V,E)$ with nodes $V = \{ 1,\dots,N\}$ and edges $E \subseteq V \times V$ is said to be  connected if there exists a path between two arbitrary nodes. $\mathcal{G}$ is said to have vertex-connectivity of $c$ if we have to cut at least $c$ nodes in $V$ to make $\mathcal{G}$ disconnected. 
\section{Distributed Decision for Interconnection of Networked Microgrids}
\label{system-problem}

This section presents a cyber-physical architecture that can support the distributed decision for electrical interconnection of multiple microgrids, and formulates  the resilient decision making problem for interconnection.

\subsection{Physical and cyber architecture of networked microgrids}
The physical-cyber architecture of microgrids is presented in Fig. \ref{fig.framework} and includes the following components:

\begin{figure}[t!]
\centering
\includegraphics[width = 3.5in]{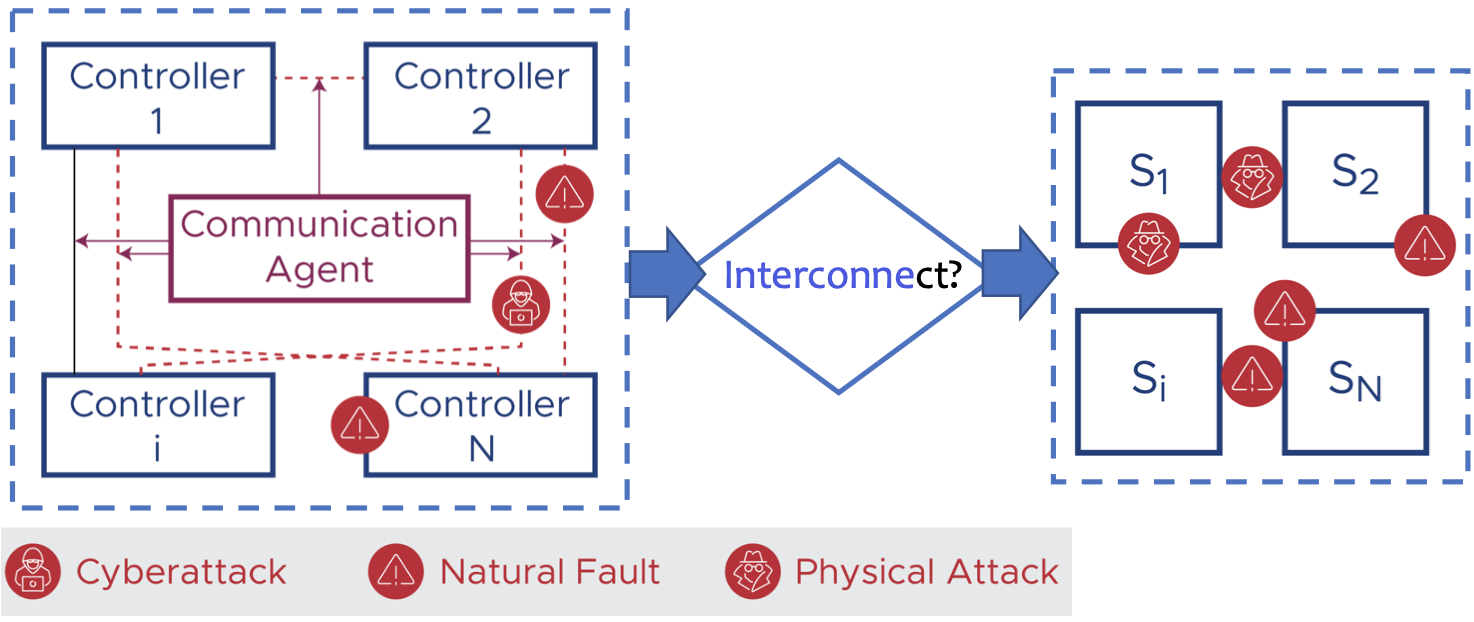}
\captionsetup{justification=raggedright,singlelinecheck=false}
\caption{\small{Physical and cyber architecture to support distributed decision making of networked microgrids. The communication agent chooses suitable pairs of microgrid-level controllers to exchange information. Using consensus algorithm, the controllers distributedly process exchanged information, even in the presence of cyberattacks, in order to reach the agreement on interconnecting the network of microgrids.}} \label{fig.framework}
\end{figure}

\begin{itemize}
    \item {\bf Physical layer.} A number of microgrids ${\bf S_i}, i=1,...,N,$ in grid-disconnected mode is considered to support the end-use demands when there is loss of the bulk power system. In this severe condition, microgrids can choose to support only critical demand. 
    
    
    \item {\bf Microgrid controllers.} For microgrids operating in a mixed ownership environment, it is impractical for one centralized controller to have access to information from all microgrids. It is more reasonable to assume that each microgrid has one controller that can collect information from devices in that microgrid and perform different controlling objectives. In this paper, the microgrid controllers aim to reach an agreement about whether the microgrids electrically interconnect or not. To make this decision distributedly, the controllers estimate local information from their microgrids and exchange such information via a p2p communication network.

    \item {\bf Communication agent.} The p2p communication links for information exchange among controllers can be vulnerable to cyberattacks. While reducing the number of communication links can mitigate this risk, it may result in inadequate information exchange among controllers. As such, appropriately controlling the p2p communication network is necessary. In this paper, a communication agent can fully control the p2p communication network (i.e., select which controllers communicate with which controllers) to mitigate the impact of cyberattacks to links. Different from a centralized controller, the communication agent does not need to collect information from microgrids.
\end{itemize}

\subsection{Interconnection criterion and Problem Statement}
The goal of interconnection is for microgrids to distributedly
determine if they should electrically interconnect,
or potentially separate, to achieve operational objectives. While several operational objectives can be considered, supporting critical load demand in the presence of cyber and physical attacks is the focus of  resilient-by-design networked microgrids. As such, in this paper the interconnection criterion is whether the total supply of the participating microgrids can meet the total critical demand. 

The interconnection decision will be evaluated in a periodical basis $T$, typically every 1 hour. At the time $nT, n=0,1,2,...,$ each microgrid controller estimates the supply and critical load demand for period  $[nT, (n+1)T]$, which are denoted as $S_i(nT)$ and $\bar{D}_i(nT)$ for other microgrids in a distributed manner. Accordingly, all the participating microgrids will electrically interconnect if the total supply can meet the total critical demand in this period, i.e.,
\begin{align}
\label{criterion}
    \sum_{i=1}^N S_i(nT)>  \sum_{i=1}^N \bar{D}_i(nT),
\end{align}
Otherwise, the microgrids operate independently. 



Therefore, the problem statement is given as follows: 

\noindent \emph{Resilient Interconnection Problem: In the presence of cyberattacks,  generate communication graph and design distributed computation scheme to enable the  microgrid controllers to evaluate the interconnection criterion \eqref{criterion} and reach the interconnection agreement.}

\section{Algorithms for Distributed Decision}
\label{algorithms}

This section presents algorithms for solving the { Interconnection problem} in Subsection II.C, including resilient consensus algorithms for distributed computation at controllers and graph algorithms for selecting suitable pairs of communicating controllers.

\subsection{Resilient Consensus Algorithm for Controllers}

Consensus algorithm is used to allow each controller to estimate  $\sum_i S_i(nT)$ and $\sum_i \bar{D}_i(nT)$ from exchanged information, however using a distributed peer-to-peer scheme. The following computation is presented for supply, but also applied for critical load demand. 
When the microgrid controller $i$ is not under any attack, it updates the supply value by using consensus algorithm as follows:
\begin{align}
    S_i^{k+1}(nT) = w_{ii} S_i^{k}(nT) + \sum_{j \in \mathcal{N}_i} w_{ij} S_j^{k}(nT), 
\end{align}
where $w_{ij}$ are the weights, $k$ is the updating step, and $\mathcal{N}_i$ is the set of all controllers that exchange supply information with controller $i$. When controller $i$ gets attacked, the attacker can inject malicious data into the updating process as follows:
\begin{align}\hspace{-.4 cm}
\label{wrongupdate}
    S_i^{k+1}(nT) = w_{ii} S_i^{k}(nT) + \sum_{j \in \mathcal{N}_i} w_{ij} S_j^{k}(nT) + u_i^k(nT), 
\end{align}
where $u_i^k(T)$ denotes data injected by the attacker.
Denote ${\bf S}^k(nT)$ as the vector of supplies and ${\bf W}$ as the matrix of weights. Then, the update \eqref{wrongupdate} can be compactly rewritten as, 
$
    {\bf S}^{k+1}(nT) = {\bf W} {\bf S}^k(nT) + {\bf B}_{\mathcal{F}} {\bf u}^k_{\mathcal{F}}(nT).
$
Here, ${\bf B}_{\mathcal{F}} = \begin{bmatrix} {\bf e}_{i1,N} & \dots & {\bf e}_{if,N} \end{bmatrix}$, where ${\bf e}_{i1,N}$ denotes an ones-vector of size $N$ with $1$ at the $i1^{th}$ row position, and ${\bf u}^k_{\mathcal{F}}(nT) = \begin{bmatrix}{\bf u}^k_{i1}(nT)^\top & \dots & {\bf u}^k_{if}(nT)^\top \end{bmatrix}^{\top}$. 

As controllers exchange supply information, controller $i^{th}$ obtains a vector of supply values as
$
    {\bf y}_i^{k}(nT) = {\bf C}_i {\bf S}^{k}(nT),
$
where ${\bf C}_i$ is a matrix with a single $1$ in each row.
After $K$ updating steps, controller $i$ has a matrix of supply values:
\begin{align}
    \label{obs}
    \mathbf{Y}_{i}^{0: K}(nT)=\mathcal{O}_{i, K} \mathbf{S}^{0}(nT)+\mathcal{M}_{i, K}^{\mathcal{F}} \mathbf{u}_{\mathcal{F}}^{0: K-1}(nT),
\end{align}
where \footnotesize $\mathbf{Y}_{i}^{0: K}(nT)=\left[\begin{array}{llll} \mathbf{y}_{i}^{ 0 \top}(nT) & \mathbf{y}_{i}^{1 \top}(nT)   & \cdots & \mathbf{y}_{i}^{K \top}(nT) \end{array} \right]^{\top}$,\\ 
$\mathbf{u}_{\mathcal{F}}^{0: K-1}(nT)= \left[\begin{array}{llll}\mathbf{u}_{\mathcal{F}}^{0 \top} & \mathbf{u}_{\mathcal{F}}^{1 \top} & \cdots & \mathbf{u}_{\mathcal{F}}^{ (K-1) \top} \end{array}\right]^{\top}$. \normalsize Also, 
\begin{align*}
    \mathcal{O}_{i, L}=\left[\begin{array}{c}
\mathbf{C}_{i} \\
\mathcal{O}_{i, L-1} \mathbf{W}
\end{array}\right], \mathcal{M}_{i, L}^{\mathcal{F}}=\left[\begin{array}{cc}
\mathbf{0} & \mathbf{0} \\
\mathcal{O}_{i, L-1} \mathbf{B}_{\mathcal{F}} & \mathcal{M}_{i, L-1}^{\mathcal{F}}
\end{array}\right], 
\end{align*}
for $L=1,2,....$ Here, $\mathcal{O}_{i, 0} = {\bf C}_i$, and $\mathcal{M}_{i, 0}^{\mathcal{F}}$ is an empty matrix with zero columns. 
Using \eqref{obs}, controller $i$ will estimate the original supplies of all microgrids. The following core result  summarizes the required connectivity condition of the communication graph for the robust estimation of the supply in the presence of a number of malicious controllers.

\begin{lemma}\cite{sundaram2010distributed}
\textit{Let the graph $\mathcal{G} = (V,E)$ with $N$ nodes denote the communication graph of controllers. Let $f$ be the maximum number of nodes that got malicious data injection during the updates. Then, each node can recover the original supply of all the other microgrids if the vertex-connectivity of the communication graph is $\geq 2f+1$ and there exist an integer $K$, and weight matrix $\bf W$ such that for set $\mathcal{Y} \subset X$ of $2f$ nodes we have, $\mbox{rank}([\mathcal{O}_{i, K} \;\;\; \mathcal{M}_{i, K}^{\mathcal{Y}}]) = N + \mbox{rank}(\mathcal{M}_{i, K}^{\mathcal{Y}})$. Moreover, $\bf W$ can be constructed with almost any choice of real-valued numbers with $w_{ij} = 0 \;\;\;\mbox{if} \;\;\;j \not\in \mathcal{N}_i \cup \{i\}$.}
\end{lemma}

When the rank condition in Lemma 1 is met, Algorithm 1 is used to estimate the total supply of all participating microgrids. 

\begin{algorithm}[H]
\KwData{Communication graph with $(2f+1)$ vertex-connectivity, and identified $f$ microgrid controllers under attack}
\KwResult{Estimates of the total supply of all participating microgrids}
 \For{n=0, 1,...}{
  \For{i=1,..., N}{
  1. \textit{Run} linear iterative updates as in \eqref{wrongupdate}.\\
  2. \textit{Get} the supply values of the neighbouring microgrids and construct the matrix of supply values $ {\bf Y}_i^{0:K}(nT)$ in \eqref{obs}. \\
  3. \textit{Construct} matrices $\mathcal{O}_{i, L}$ and $\mathcal{M}_{i, L}^{\mathcal{F}}$.\\
  }
  4. \textit{Estimate} ${\bf S}^{0}(nT)$ using the inverse of \eqref{obs} \\
  5. \textit{Calculate} total supply as $\sum_{i=1}^N {\bf S}^{0}(nT).$
 }
 \caption{Estimation of total supply}
\end{algorithm}

\subsection{Graph algorithms for communication agent}
\label{sec.communicationagent}
\textcolor{black}{As seen in Lemma 1, the key to the success of resilient consensus algorithm is to have a communication graph with vertex-connectivity of at least $(2f+1)$ if there are $f$ attacked controllers, while the weight matrix $\bf W$ can be chosen arbitrarily. However, to the best of our knowledge, there is no algorithms to generate a graph with a given vertex-connectivity.} 
In this work, we develop the  following core result and use it  to generate such communication graphs.


\begin{lemma} [\bf  Graph Extension Scheme]
\label{graphlemma}
Let $\mathcal{G}_n$ be a graph with $n$ nodes and vertex-connectivity $m$. Extend the graph $\mathcal{G}_n$ to a $(n+1)$-node graph $\mathcal{G}_{(n+1)}$ by adding 1 node to graph $\mathcal{G}_n$ and $m$ edges from this added node to $m$ arbitrary nodes of $\mathcal{G}_n$. Then the new graph $\mathcal{G}_{(n+1)}$ has vertex-connectivity larger than or equal to $m$.
\end{lemma}

{\bf Proof:} See Appendix \ref{proof}.

In the following, two scenarios of cyberattacks to communication links are discussed, followed by algorithms for communication agent to generate communication network.

\subsubsection{Preventive control}
 To reduce the likelihood and impact of successful attacks to the communication links, we propose a randomization strategy, whereby  the communication agent  generates the p2p communication graph with $(2f+1)$-vertex-connectivity by randomly selecting the nodes and ordering them.


The preventive strategy is provided in Algorithm 2.  
Note  that the $(2f+1)$-node fully connected graph generated in Step 2 has the vertex-connectivity of $(2f+1)$. As the graph is subsequently extended in Step 3, the vertex-connectivity of the graph is larger than or equal to $(2f+1)$, as proved in Lemma \ref{graphlemma}. Therefore, the $N$-node graph obtained by Step 3 also has the vertex-connectivity  larger than or equal to $(2f+1)$. The randomization of node order in Step 1 and Step 4 does not change the vertex-connectivity of the graph. Hence, Algorithm 2 results in a $N$-node graph with vertex-connectivity larger than or equal to $(2f+1)$.

\begin{algorithm}[H]
\KwData{Number of microgrids (i.e., nodes), $N$}
\KwResult{A random p2p graph with $(2f+1)$-vertex-connectivity for preventive control}
 1. Randomly select $(2f+1)$ nodes among $N$ nodes.

 2. Generate fully connected graph from these $(2f+1)$ nodes 

 3. Subsequently extend the graph using the Graph Extension Scheme presented in Lemma 2 until $N$-node graph is obtained.

 4. Randomly change the node order  of the resulting graph from Step 3 
\caption{P2p graph for preventive control}
\end{algorithm}

\subsubsection{Responsive control }

When there are cyberattacks to communication links and they are known, the communication agent will adapt the network by generating p2p communication network with $(2f+1)$-vertex-connectivity from the remaining safe communication links. Here we assume the ratio of compromised communication links  is sufficiently small to ensure the $(2f+1)$-vertex-connectivity requirement is met. The procedure  is presented in Algorithm 3.

\begin{algorithm}[H]
\KwData{Number of microgrids (nodes), $N$}
\KwResult{A p2p graph with $(2f+1)$-vertex-connectivity for responsive control}
 1.  Select $(2f+1)$ nodes among  nodes whose communication links do not get attacked.

 2. Generate fully connected graph from the $(2f+1)$ nodes obtained in Step 1 

 3. Subsequently add the safe communication links to extend the graph using the Graph Extension Scheme presented in Lemma 2 until  $N$-node graph is obtained.

\caption{P2p graph for responsive control}
\end{algorithm}

It is noted that in Step 1, only the controller nodes with healthy (no attack) communication links  are selected. Therefore, the resulting fully connected graph in Step 2 consists of non-attacked communication links. In Step 3, only the safe communication links are used to extend the graph. Therefore, Algorithm 3 results in a p2p communication graph with safe communication links (edges) and vertex-connectivity  larger than or equal to $(2f+1)$.

\section{Demonstration}
\label{demonstration}

\begin{figure}
    \centering
    \includegraphics[width=1.05\columnwidth]{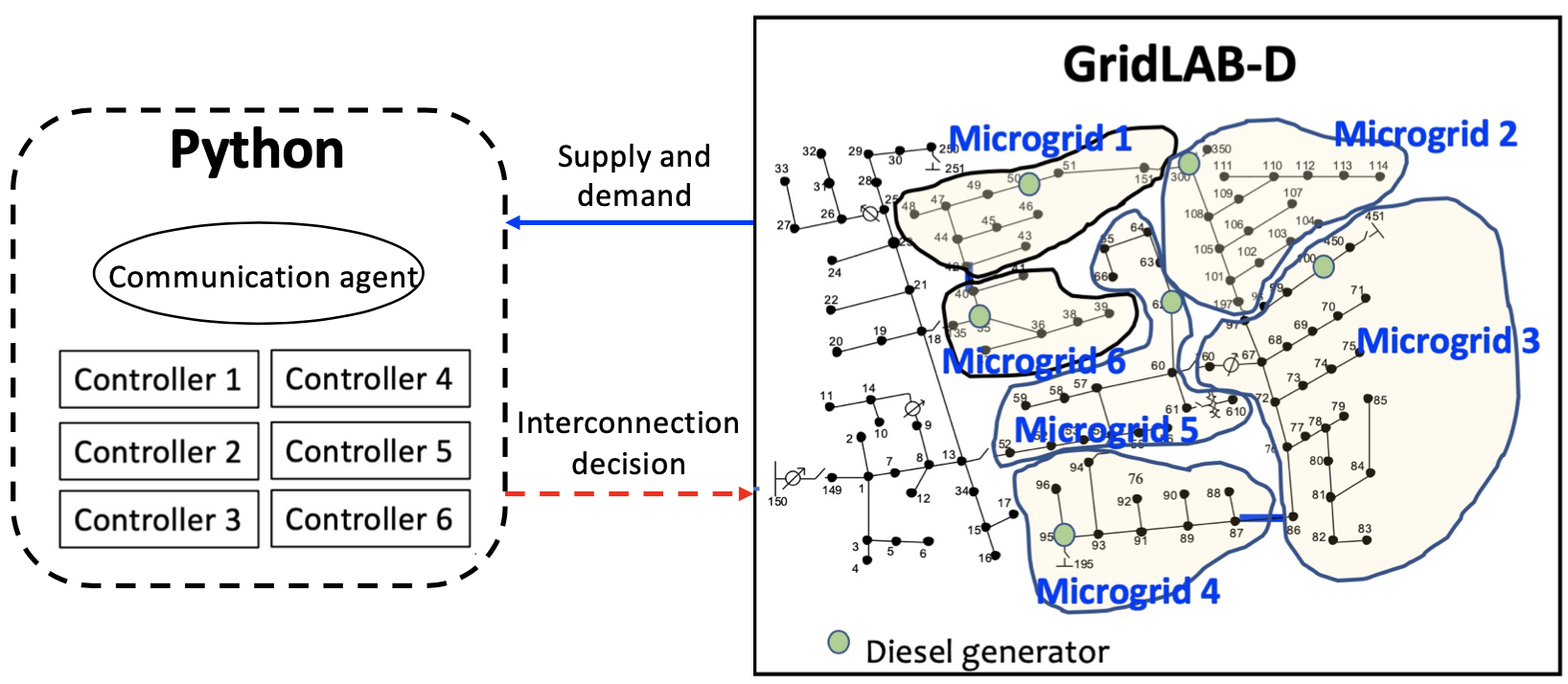}
    \vspace{-2mm}
    \captionsetup{justification=raggedright,singlelinecheck=false}
	\caption{\small{Co-simulation platform for demonstrating the resilient communication scheme. Microgrid controllers and communication agent are programmed in Python. Microgrid controllers distributedly decide if microgrids should interconnect. It should be noted that after the interconnection decision is reached, then microgrids can electrically interconnect using available switches, and the critical demand will be met by the supply in any electrical interconnection topology. How to optimize the electrical interconnection topology is beyond the scope of this paper.}}
	\vspace{-2mm}
	 \label{fig.co-sim}
\end{figure}

For demonstrating the introduced resilient communication scheme, a co-simulation platform is developed, as showed in Fig. \ref{fig.co-sim}, in which the microgrids are modeled in GridLAB-D \cite{chassin_gridlab-d_2014} and microgrid controllers are programmed in Python. 
A modified IEEE 123-node test feeder is divided into six microgrids, each of which contains one generator.

The locations of generators and critical loads are in Table \ref{tab:location}. At a particular  time, the Python-based controllers get measurement data from GridLAB-D and estimate the supply and critical demands of microgrids for the next 1 hour, as in Table \ref{tab:location}. 
It can be observed that the supplies at MGs 3 and 5 do not meet the critical load demand, and hence, it is beneficial to interconnect the MGs.

\begin{table}[]
\renewcommand{\arraystretch}{1.2}
\caption{Local information of microgrids}
\centering
\begin{tabular}{c c c c c c}
\label{tab:location}
MG & Gen   & Critical load  & Critical demand & Supply    \\
   & nodes & nodes          & (kVA-h)  &(kVA-h) \\
\hline
1        &  50&          (46) & 22.30            & 24.17 \\
2        &     300&         (111-114)  & 44.72               & 64.31\\
3        &      100 &     (68-71) & 111.80            & 89.19  \\
4        &       95&     (88,94)&  89.44           & 134.43 \\
5        &        62&    (52,53,55) &  89.44             & 49.65  \\
6        &      35&        (39) & 22.36              & 79.69  \\
\hline
\end{tabular}
\end{table}

\begin{figure}
    \centering
    \includegraphics[width=0.6\columnwidth]{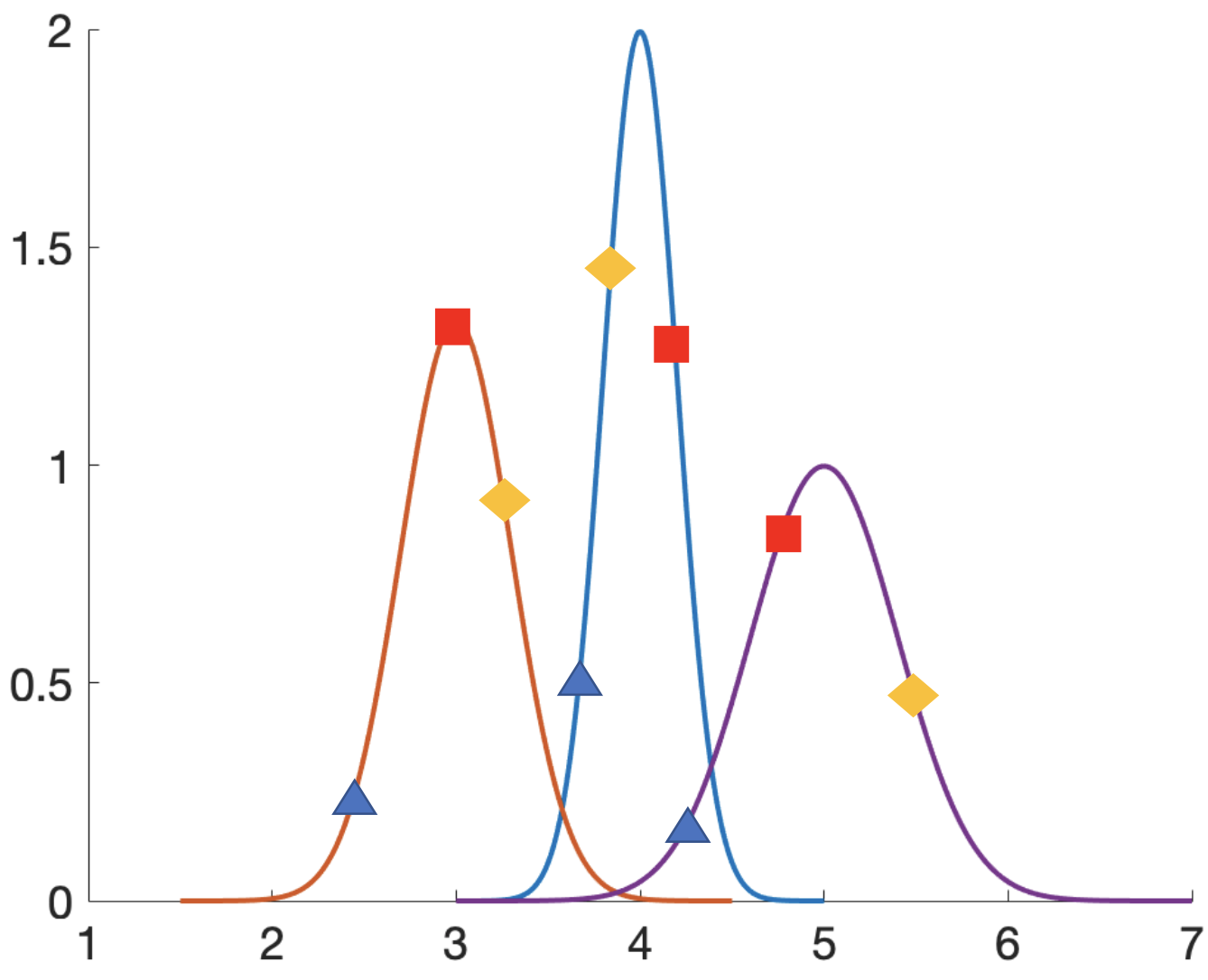}
    \vspace{-2mm}
    \captionsetup{justification=raggedright,singlelinecheck=false}
	\caption{\small{Distributions of noises for the attacker to sample data to inject into the computation software at MG4's controller. In each distribution, three samples are arbitrarily selected.}}
	\vspace{-2mm}
	 \label{fig.noise}
\end{figure}
\begin{figure}[]
    \centering
    \includegraphics[width = .7\linewidth]{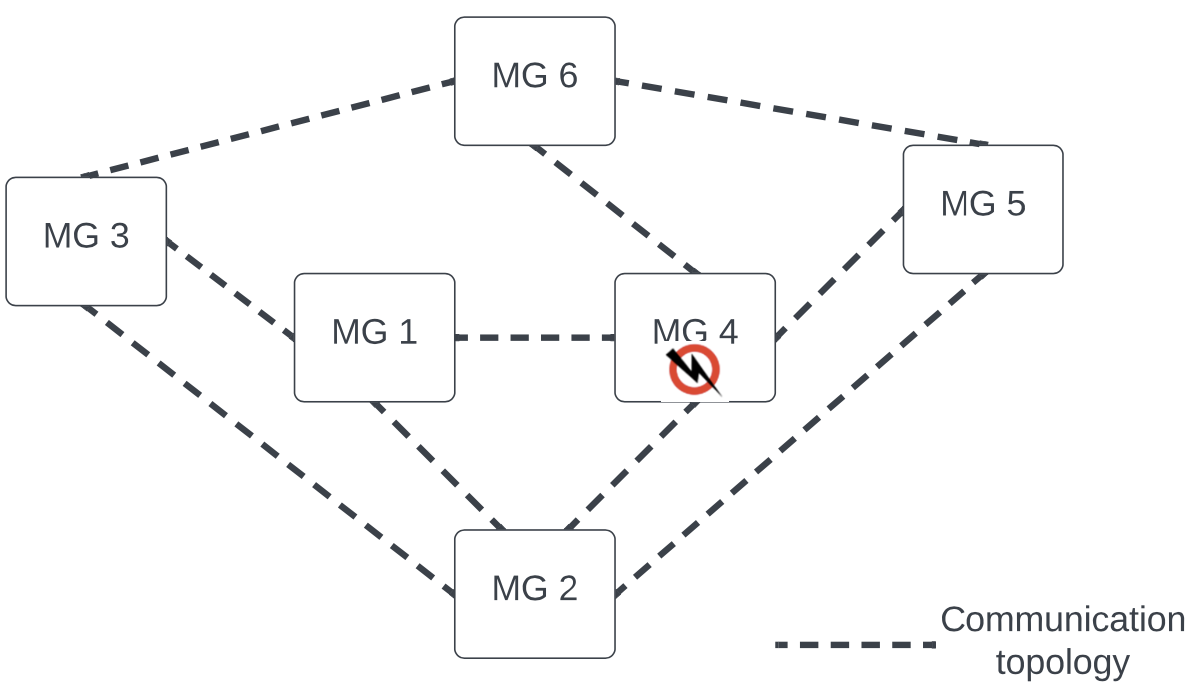}
    \captionsetup{justification=raggedright,singlelinecheck=false}
    \caption{\small{Random communication topology with 3-vertex-connectivity. MG $4$ malfunctions.}}
    \label{fig:microgrids1}
\end{figure}

Assume that all the microgrid controllers choose to participate in the interconnection decision process and MG4's controller is under attack. The attacker injects malicious data into the decision making algorithm for three consecutive update steps, and therefore, the supply and demand exchanged between the controllers are inaccurate. The malicious data is arbitrarily sampled from 3 distributions, as presented in Fig. \ref{fig.noise}. As there is 1 controller that got attacked, communication agent generates a communication graph with 3-vertex-connectivity by using Algorithm 2, as shown in  Fig. \ref{fig:microgrids1}.


Next,  Algorithm $1$ is used to estimate the total supply and critical load demand values of all the microgrids. The weight matrix is selected as:
\begin{align}
\label{Wmatrix}
    {\bf W} = \begin{bmatrix}5  &   4   &  1  &   1   &  0  &   0\\
     1  &  -3  &  -2  &   1  &   4  &   0\\
    -3  &  -3  &  -4  &   0   &  0  &  -3\\
    -1  &  -2  &   0  &   5 &   -1 &   -3\\
     0  &  4  &   0  &   5  &  -1 &   -4\\
     0  &   0  &  -3  &  -1 &    1  &  -3
 \end{bmatrix}.
\end{align}

\begin{figure}
\begin{subfigure}{0.5\columnwidth}
  \centering
  \includegraphics[width=.95\columnwidth]{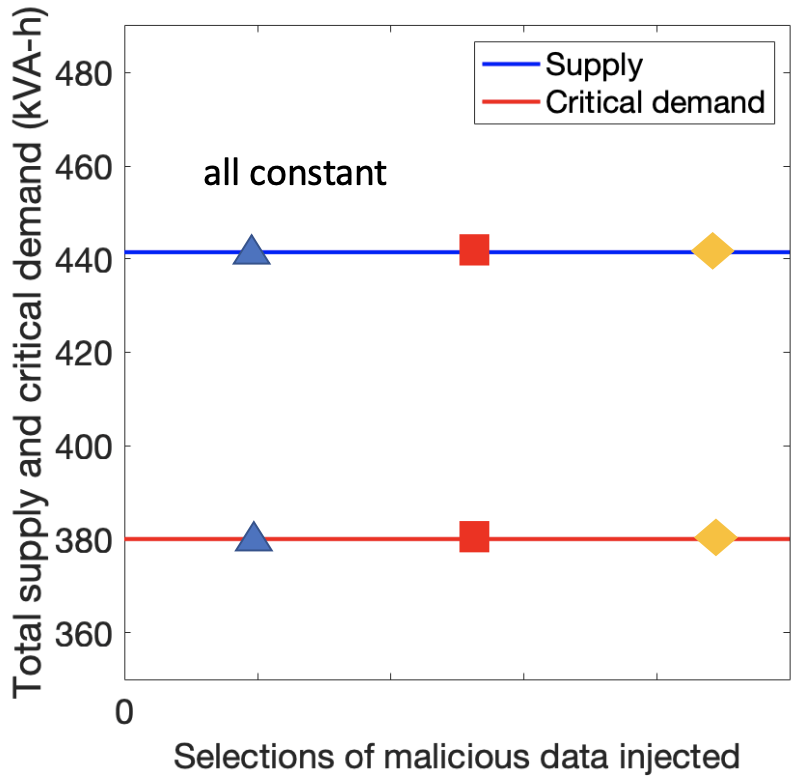}
  \caption{Resilient consensus algorithm}
  \label{fig:sfig1}
\end{subfigure}%
\begin{subfigure}{0.5\columnwidth}
  \centering
  \includegraphics[width=.95\columnwidth]{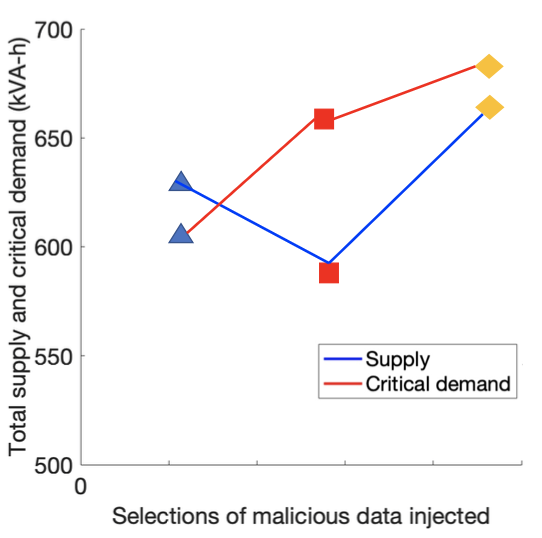}
  \caption{Baseline}
  \label{fig:sfig2}
\end{subfigure}
\caption{Comparison between resilient consensus algorithm and baseline performance under malicious data attack}
\label{fig:microgrids1_est}
\end{figure}

The average consensus algorithm is used as the baseline for comparison with the resilient consensus algorithm. The performance of two algorithms are showed in Fig. \ref{fig:microgrids1_est}. It can be observed that using resilient consensus algorithm, each microgrid controller can accurately estimate the total supply and demand, in the presence of the data injection attack, as: $\sum_{i=1}^6 S = 441.36$ (kVA-h), $\sum_{i=1}^6 \bar{D}= 380.04$ (kVA-h). As such, all the microgrid reached a consensus that the interconnection criterion \eqref{criterion} is satisfied. But, if controllers use the baseline algorithm, they cannot accurately estimate the total supply and demand, and they cannot evaluate interconnection criterion \eqref{criterion}.


\section{Conclusions}
In this paper,  a resilient communication scheme was introduced for distributed decision of interconnecting networks of microgrids. Algorithms were presented to generate a randomized  p2p communication network to prevent or limit a successful attack and an adapted network for situations where some of the  communication links are attacked and known. Based on the p2p communication network, resilient consensus algorithm was applied to enable the microgrid controllers to estimate global information of total supply and critical demand, and then, agree to physically interconnect or disconnect the network, even in the presence of cyberattacks. Simulations on a modification of the 123-node Test Feeder with 6 microgrids showed that the controllers can distributedly make  decision under a data injection attack.

\section{Appendix: Proof of Lemma \ref{graphlemma}}
\label{proof}
Let's remove $(m-1)$ arbitrary nodes and all the edges connected to these nodes from the graph $\mathcal{G}_{(n+1)}$. Denote the remaining graph as $\tilde{G}_{(n+1)}.$
There are two possibilities:

Case 1: The added node to extend from the graph $\mathcal{G}_{n}$ to graph $\mathcal{G}_{(n+1)}$ is among the (m-1) removed nodes. Now, the remaining graph $\tilde{G}_{(n+1)}$ is actually the one obtained by removing $(m-2)$ nodes from the graph $\mathcal{G}_{n}$. As the graph $\mathcal{G}_{n}$ is $m$-vertex-connectivity, it is still connected after cutting $(m-2)$ nodes. Hence, the graph $\tilde{G}_{(n+1)}$ is connected.

Case 2:  The added node to extend from the graph $\mathcal{G}_{n}$ to graph $\mathcal{G}_{(n+1)}$ is not removed. Then, the remaining graph $\tilde{G}_{(n+1)}$ includes the added node and a graph obtained by removing $(m-1)$ nodes from the graph $\mathcal{G}_{n}$ (which is denoted as $\tilde{G}_{n}$). Because the graph $\mathcal{G}_{n}$ is $m$-vertex-connectivity, the graph $\tilde{G}_{n}$ is also connected. Since the added node is connected to m nodes in the graph $\mathcal{G}_{n}$ due to the Graph Extension Scheme, it should be connected to at least one node in the graph  $\tilde{G}_{n}$. Hence,  $\tilde{G}_{(n+1)}$ is connected.

Therefore, when we remove any (m-1) nodes from the graph $\mathcal{G}_{(n+1)}$, then the remaining graph is still connected. Hence,  $\mathcal{G}_{(n+1)}$ has the vertex-connectivity of at least $m$.

\bibliographystyle{IEEEtran}
\bibliography{ref}

\begin{thebibliography}{10}
\providecommand{\url}[1]{#1}
\csname url@samestyle\endcsname
\providecommand{\newblock}{\relax}
\providecommand{\bibinfo}[2]{#2}
\providecommand{\BIBentrySTDinterwordspacing}{\spaceskip=0pt\relax}
\providecommand{\BIBentryALTinterwordstretchfactor}{4}
\providecommand{\BIBentryALTinterwordspacing}{\spaceskip=\fontdimen2\font plus
\BIBentryALTinterwordstretchfactor\fontdimen3\font minus
  \fontdimen4\font\relax}
\providecommand{\BIBforeignlanguage}[2]{{%
\expandafter\ifx\csname l@#1\endcsname\relax
\typeout{** WARNING: IEEEtran.bst: No hyphenation pattern has been}%
\typeout{** loaded for the language `#1'. Using the pattern for}%
\typeout{** the default language instead.}%
\else
\language=\csname l@#1\endcsname
\fi
#2}}
\providecommand{\BIBdecl}{\relax}
\BIBdecl

\bibitem{7922501}
Z.~Li, M.~Shahidehpour, F.~Aminifar, A.~Alabdulwahab, and Y.~Al-Turki,
  ``Networked microgrids for enhancing the power system resilience,''
  \emph{Proc. IEEE}, vol. 105, no.~7, pp. 1289--1310, 2017.

\bibitem{yuan2013resilient}
Y.~Yuan, Q.~Zhu, F.~Sun, Q.~Wang, and T.~Ba{\c{s}}ar, ``Resilient control of
  cyber-physical systems against denial-of-service attacks,'' in \emph{2013 6th
  International Symposium on Resilient Control Systems (ISRCS)}.\hskip 1em plus
  0.5em minus 0.4em\relax IEEE, 2013, pp. 54--59.

\bibitem{bai2017data}
C.-Z. Bai, F.~Pasqualetti, and V.~Gupta, ``Data-injection attacks in stochastic
  control systems: Detectability and performance tradeoffs,''
  \emph{Automatica}, vol.~82, pp. 251--260, 2017.

\bibitem{zhu2013performance}
M.~Zhu and S.~Martinez, ``On the performance analysis of resilient networked
  control systems under replay attacks,'' \emph{IEEE Transactions on Automatic
  Control}, vol.~59, no.~3, pp. 804--808, 2013.

\bibitem{barboni2020detection}
A.~Barboni, H.~Rezaee, F.~Boem, and T.~Parisini, ``Detection of covert
  cyber-attacks in interconnected systems: A distributed model-based
  approach,'' \emph{IEEE Transactions on Automatic Control}, 2020.

\bibitem{8618618}
K.~P. Schneider, S.~Laval, J.~Hansen, R.~B. Melton, L.~Ponder, L.~Fox, J.~Hart,
  J.~Hambrick, M.~Buckner, M.~Baggu, K.~Prabakar, M.~Manjrekar, S.~Essakiappan,
  L.~M. Tolbert, Y.~Liu, J.~Dong, L.~Zhu, A.~Smallwood, A.~Jayantilal,
  C.~Irwin, and G.~Yuan, ``A distributed power system control architecture for
  improved distribution system resiliency,'' \emph{IEEE Access}, vol.~7, pp.
  9957--9970, 2019.

\bibitem{selfhealingNM}
Z.~Wang, B.~Chen, J.~Wang, and C.~Chen, ``Networked microgrids for self-healing
  power systems,'' \emph{IEEE Transactions on Smart Grid}, vol.~7, no.~1, pp.
  310--319, 2016.

\bibitem{schneider2022framework}
K.~P. Schneider, J.~Glass, C.~Klauber, B.~Ollis, M.~J. Reno, M.~Burck,
  L.~Muhidin, A.~Dubey, W.~Du, L.~Vu \emph{et~al.}, ``A framework for
  coordinated self-assembly of networked microgrids using consensus
  algorithms,'' \emph{IEEE Access}, 2022.

\bibitem{sundaram2010distributed}
S.~Sundaram and C.~N. Hadjicostis, ``Distributed function calculation via
  linear iterative strategies in the presence of malicious agents,'' \emph{IEEE
  Trans. Automatic Control}, vol.~56, no.~7, pp. 1495--1508, 2010.

\bibitem{chassin_gridlab-d_2014}
D.~P. Chassin, J.~C. Fuller, and N.~Djilali,
  ``\BIBforeignlanguage{en}{{GridLAB}-{D}: {An} {Agent}-{Based} {Simulation}
  {Framework} for {Smart} {Grids}},'' \emph{\BIBforeignlanguage{en}{Journal of
  Applied Mathematics}}, vol. 2014, p. e492320, Jun. 2014.

\end{thebibliography}
\end{document}